\documentclass[12pt,a4paper]{article}
\usepackage[english]{babel}
\usepackage[latin1]{inputenc}
\usepackage[T1]{fontenc}
\usepackage{amsmath}
\usepackage{amsfonts}
\usepackage{amssymb}
\usepackage{braket}
\usepackage[dvips]{graphicx,color}
\usepackage{hyperref}
\begin{document}
\title{Dynamical Casimir effect for fermions in $2+1$ dimensions} 
\author{C.~D.~Fosco and
G.~Hansen\\
{\normalsize\it Centro At\'omico Bariloche and Instituto Balseiro}\\
{\normalsize\it Comisi\'on Nacional de Energ\'{\i}a At\'omica}\\
{\normalsize\it R8402AGP S.\ C.\ de Bariloche, Argentina.} }
%%%%%%%%%%%%%%%%%%%%%%%%%%%%%%%%%%%%%%%%%%%%%%%%%%%%%%%%%%%%%%%%%%%%%%%%%%%%%%
%%%%%%%%%%%%%%%%%%%%%%%%%%%%%%%%%%%%%%%%%%%%%%%%%%%%%%%%%%%%%%%%%%%%%%%%%%%%%%
%%%%%%%%%%%%%%%%%%%%%%%%%%%%%%%%%%%%%%%%%%%%%%%%%%%%%%%%%%%%%%%%%%%%%%%%%%%%%%
%%%%%%%%%%%%%%%%%%%%%%%%%%%%%%%%%%%%%%%%%%%%%%%%%%%%%%%%%%%%%%%%%%%%%%%%%%%%%%
%%%%%%%%%%%%%%%%%%%%%%%%%%%%%%%%%%%%%%%%%%%%%%%%%%%%%%%%%%%%%%%%%%%%%%%%%%%%%%
%\date{}
\maketitle
%========================================================================
\begin{abstract} 
We study the fermion pair creation phenomenon due to the time dependence
of curves, where boundary conditions are imposed on a Dirac field in
$2+1$ dimensions. These conditions, which lead to non-trivial relations
for the normal component of the fermionic current, depend on the value
of a dimensionless parameter. We show that the pair creation effect is
maximized for bag boundary conditions, obtained for a particular value
of that parameter. 
The effect is studied in terms of the effective action to extract
information on the probability of vacuum decay, using an expansion
in powers of the deformation of the curves with respect to straight lines.
We demonstrate that the first non-trivial contributions to this process
can  be obtained from the electromagnetic vacuum polarization tensor for
a Dirac field coupled to {\em static\/} boundaries.
\end{abstract}
%=============================================================================
\maketitle
%=============================================================================
%%%%%%%%%%%%%%%%%%%%%%%%%%%%%%%%%%%%%%%%%%%%%%%%%%%%%%%%%%%%%%%%%%%%%%%%%%%%%%
%%%%%%%%%%%%%%%%%%%%%%%%%%%%%%%%%%%%%%%%%%%%%%%%%%%%%%%%%%%%%%%%%%%%%%%%%%%%%%
%%%%%%%%%%%%%%%%%%%%%%%%%%%%%% Introduction %%%%%%%%%%%%%%%%%%%%%%%%%%%%%%%%%%
%%%%%%%%%%%%%%%%%%%%%%%%%%%%%%%%%%%%%%%%%%%%%%%%%%%%%%%%%%%%%%%%%%%%%%%%%%%%%%
%%%%%%%%%%%%%%%%%%%%%%%%%%%%%%%%%%%%%%%%%%%%%%%%%%%%%%%%%%%%%%%%%%%%%%%%%%%%%%
\section{Introduction}
Many interesting Quantum Field Theory effects arise from the presence of
boundaries affecting one or more of the fields relevant to the phenomenon
being described. A well-known example of this is the Casimir
effect~\cite{Casimir:1948,Bordag:2001} which, in its static version,
manifests itself in the existence of macroscopic forces between neutral
objects.  These forces exhibit a non-trivial dependence of the vacuum
energy on the geometry of the boundaries.

Non-static boundaries, the case that we consider here, lead instead to the
dynamical Casimir effect (DCE), whereby a suitable time dependence in the
boundary conditions leads to the creation of particles, real quanta of the
vacuum field. It should be noted that the time-dependence of the boundary
conditions may be due to a rapid variation of the media properties, without
necessarily changing their geometry.
Regardless of its specific origin, the situation is that of an open system.
Typically, it is necessary for an agent to exert work on the system, which is
then converted into radiation: a dissipative effect. This effect depends,
among other things, on the kind of vacuum field involved. The most
extensively studied case is that of an Abelian gauge field in $3+1$
dimensions, not just because of the ubiquitous nature of the
electromagnetic (EM) field, but also due to the fact that the required
boundary conditions can be experimentally imposed in a rather
straightforward fashion.  In this paper, we deal with the Dirac field in
$2+1$ space-time dimensions, a kind of field that finds a realization in
the realm of Condensed Matter Physics~\cite{Fradkin:2013}. In the same
context, boundary conditions may also be relevant, due to the existence of
impurities, domain walls, and other kinds of defects.  In this and related
contexts, interesting  works continue to appear; for instance dealing with
the Casimir effect the lattice
fermions~\cite{Ishikawa:2020,Mandlecha:2022}, or in the presence of
explicit Lorentz symmetry breaking~\cite{Cruz:2018}.

In this paper, we study a system consisting of Dirac fermions in $2+1$
dimensions, coupled to moving boundaries.  This work generalizes the
results presented in~\cite{OscBag1+1} to more than one dimension and shows
novel relations with apparently unrelated effects, like vacuum
polarization.

This paper is organized as follows: in Sect.~\ref{sec:themodel}, we
describe the model that we study in this work, and introduce our notation
and conventions. The total probability of fermion pair creation is
evaluated in Sect.~\ref{sec:eff}, via the calculation of the effective
action of the system and its imaginary part. We do that for different
situations, involving one or two walls.
Finally, in Sect.~\ref{sec:conc} we present our conclusions.

%%%%%%%%%%%%%%%%%%%%%%%%%%%%%%%%%%%%%%%%%%%%%%%%%%%%%%%%%%%%%%%%%%%%%%%%%%%%%%
%%%%%%%%%%%%%%%%%%%%%%%%%%%%%%%%%%%%%%%%%%%%%%%%%%%%%%%%%%%%%%%%%%%%%%%%%%%%%%
%%%%%%%%%%%%%%%%%%%%%%%%%%%%%%%%%%%%%%%%%%%%%%%%%%%%%%%%%%%%%%%%%%%%%%%%%%%%%%
%%%%%%%%%%%%%%%%%%%%%%%%%%%%%%% The model %%%%%%%%%%%%%%%%%%%%%%%%%%%%%%%%%%%%
%%%%%%%%%%%%%%%%%%%%%%%%%%%%%%%%%%%%%%%%%%%%%%%%%%%%%%%%%%%%%%%%%%%%%%%%%%%%%%
%%%%%%%%%%%%%%%%%%%%%%%%%%%%%%%%%%%%%%%%%%%%%%%%%%%%%%%%%%%%%%%%%%%%%%%%%%%%%%
%%%%%%%%%%%%%%%%%%%%%%%%%%%%%%%%%%%%%%%%%%%%%%%%%%%%%%%%%%%%%%%%%%%%%%%%%%%%%%
\section{The model}\label{sec:themodel}
We shall work with a model which we conveniently define by its (real-time)
action ${\mathcal S}$, a functional of a Dirac field $\psi$ and its adjoint
 ${\bar\psi}$, as well as of an external function $V$: 
\begin{equation}\label{eq:defsf}
{\mathcal S}({\bar\psi},\psi;V)\;=\;
\int d^3x \,{\bar\psi}(x) \, {\mathcal D} \, \psi(x) \;,
\end{equation}
where the operator ${\mathcal D}$ is given by  
\begin{equation}
{\mathcal D} \equiv i\not \! \partial - m - V(x) \;.
\end{equation}
The ``potential'' $V(x)$ is used to introduce boundary conditions (see
below), and $m$ denotes the mass of the fermion field. In our conventions,
both $\hbar$ and the speed of light are set equal to $1$, the space-time
coordinates are denoted by $x^\mu$, $\mu\,=\, 0,\,1,\,2$, $x^0 = t$, and we
use the Minkowski metric \mbox{$g_{\mu\nu} \equiv {\rm diag}(1,-1,-1)$}.
Dirac's $\gamma$-matrices, on the other hand, are chosen to be in the
representation: $\gamma^0 \equiv \sigma_1$,  $\gamma^1 \equiv i \sigma_2$, 
$\gamma^2 \equiv i \sigma_3$,  where:
\begin{equation}\label{eq:gamma_matrices}
\sigma_1 \,=\, 
\left(
\begin{array}{cc}
	0 & 1 \\
	1 & 0
\end{array}
\right)
\;,\;\;
\sigma_2 \,=\, 
\left(
\begin{array}{cc}
	0 & -i\\
	i & 0 
\end{array}
\right) \;,
\end{equation}
and 
\begin{equation}
\sigma_3 \,=\,\left(
\begin{array}{cc}
	1 & 0 \\
	0 & -1
\end{array}
\right) \,,  
\end{equation}
where $\sigma_i$ ($i=1,\,2,\,3$) denote the usual Pauli's matrices.

Following the approach of~\cite{Fosco:2007,Fosco:2008,Ttira:2010}, to
impose boundary conditions on a given space-time region $\Sigma$, we use a
potential $V$, proportional to $\delta_\Sigma(x)$; namely, a Dirac's
$\delta$ function concentrated on a manifold of co-dimension $1$. Since we
work in $2+1$ dimensions, $\Sigma$ will be the surface(s) swept by
time-dependent spatial curve(s).
   
Let us briefly recall the kind of boundary condition introduced by this
kind of potential in a particularly simple case, \mbox{$V(x) =
g \delta(x^2)$}. We can use the following heuristic argument: integrating
the Dirac equation from $x^2 = -\epsilon$ to $x^2 = \epsilon$, we get a
possible discontinuity in the Dirac field at $x^2=0$, since its jump at
that location is related to its value at $x^2 = 0$.
Following~\cite{FoscoTorroba:2007}, we replace the integral of the
$\delta$-function times $\psi$ by the average of the two lateral limits:
\begin{equation} 
i \gamma^2 ( \psi(x_\shortparallel,\epsilon) - 
\psi(x_\shortparallel, -\epsilon) ) \,-\,
\frac{g}{2} \big( \psi(x_\shortparallel,\epsilon) +
\psi(x_\shortparallel,-\epsilon) \big)  \;=\;0 \;,
\end{equation}
where $x_\shortparallel = (x^0,x^1)$.
Introducing the orthogonal projectors: ${\mathcal P}^{\pm} \equiv \frac{1
\pm i \gamma^2}{2}$, this is equivalent to: 
\begin{equation}\label{eq:Eproyectors} 
( 1 \mp g/2) \, {\mathcal P}^{\pm}\psi(x_\shortparallel, \epsilon) \;=\; 
( 1 \pm g/2) \, {\mathcal P}^{\pm}\psi(x_\shortparallel,-\epsilon) \;.  
\end{equation} 
This implies, for $g=2$, that ${\mathcal P}^{\pm}\psi(x_\shortparallel, \mp
\epsilon) = 0$, i.e., bag boundary conditions \cite{Fosco:2008,Chodos:1974}
on both sides of the wall.
The previous formal argument will be seen to hold true in more concrete
terms, in what follows: indeed, as it was the case for the static Casimir
effect for a Dirac field, we shall see that the dissipative
effects are also maximized for the very same value: $g=2$.
 
In this work, we deal with two cases, depending on whether boundary
conditions are imposed on one or two curves. The space-time geometry swept
by each curve will be defined in terms of a single function, $\varphi$,
namely, \mbox{$x^2\,=\,\varphi(x_\shortparallel)$}.  We will use indices
from the beginning of the Greek alphabet ($\alpha,\,\beta,\, \ldots$),
taking the values $0$ and $1$, to label the coordinates of
$x_\shortparallel$, used to parametrize the boundaries.  

When dealing with two boundaries, which we will denote by $L$ and $R$, we
shall distinguish between two situations: firstly, $L$ will be regarded in
motion, and determined by the equation $x^2\,=\,\varphi_L(x_\shortparallel)$,
while the other, $R$, will be defined by $x^2\,=\, a > 0$. 
Secondly, the two boundaries will be allowed to move, such that $L$ and $R$
will be determined by $x^2\,=\, \varphi_L(x_\shortparallel)$ and
$x^2\,=\, a + \varphi_R(x_\shortparallel)$, respectively.

The form of $V$ for a single boundary shall be assumed to be given by the
expression:
\begin{equation}
\label{eq:V_onewall}
V(x) \;=\; g \, \sqrt{1 - \partial_\alpha \varphi(x_\shortparallel)
\partial^\alpha \varphi(x_\shortparallel)} \; \delta\big(x^2 -
\varphi(x_\shortparallel) \big) \;,
\end{equation}
where the square root factor has been included to ensure reparametrization
invariance of the corresponding term in the action. This is required, on
physical grounds, by the (assumed) homogeneity of the boundary conditions on
the region defined by $V$.

Finally, when two walls are present, one simply adds the corresponding
potentials. For example, when $L$ moves and $R$ is static,
\begin{equation}\label{eq:V_twowalls}
V(x) \;=\; g_L \, \sqrt{1 - \partial_\alpha \varphi_L(x_\shortparallel)
\partial^\alpha \varphi_L(x_\shortparallel)} \; \delta\big(x^2 -
\varphi_L(x_\shortparallel)\big) \,+\, g_R \, \delta(x^2 - a) \;,
\end{equation}
where the constants $g_L$ and $g_R$ can be set equal to their bag-model
values $g_L=g_R=2$ at any point in the calculation.

%=============================================================================
%%%%%%%%%%%%%%%%%%%%%%%%%%%%%%%%%%%%%%%%%%%%%%%%%%%%%%%%%%%%%%%%%%%%%%%%%%%%%%
%%%%%%%%%%%%%%%%%%%%%%%%%%%%%%%%%%%%%%%%%%%%%%%%%%%%%%%%%%%%%%%%%%%%%%%%%%%%%%
%%%%%%%%%%%%%%%%%%%%%%%%%%%% Effective action %%%%%%%%%%%%%%%%%%%%%%%%%%%%%%%%
%%%%%%%%%%%%%%%%%%%%%%%%%%%%%%%%%%%%%%%%%%%%%%%%%%%%%%%%%%%%%%%%%%%%%%%%%%%%%%
%%%%%%%%%%%%%%%%%%%%%%%%%%%%%%%%%%%%%%%%%%%%%%%%%%%%%%%%%%%%%%%%%%%%%%%%%%%%%%
\section{Imaginary part of the effective action}\label{sec:eff}
Let us introduce here the (in-out) effective action $\Gamma$, and then
evaluate the leading term contributing to its imaginary part, using a
perturbative approach, in powers of the amplitude of deviation of the
boundaries with respect to static straight lines. 

$\Gamma$ is obtained, in the functional integral formulation, by
integrating out the Dirac field,
\begin{equation}\label{eq:defgef}
e^{i \Gamma} \;=\; \frac{\int {\mathcal D}\psi {\mathcal
D}{\bar\psi} \, e^{i {\mathcal S}({\bar\psi},\psi;V)}}{\int 
{\mathcal D}\psi {\mathcal D}{\bar\psi} \, 
e^{i {\mathcal S}({\bar\psi},\psi;V_0)}} \;.
\end{equation}
Here, $V$ is assumed to be the one corresponding to the situation being
considered, namely, one or two walls, while $V_0$ is the function $V$
restricted  to  static straight lines ($\varphi = 0$, or \mbox{$\varphi_L =
\varphi_R = 0$}, depending on the case).   
The denominator thus incorporates the static fermionic Casimir
effect~\cite{Fosco:2008}, while the effective action encompasses the
strictly dynamical effects, since (we assume) the time average of the
deformation vanishes. Should this not be the case, any finite value for
that average could be compensated by a redefinition of the $x^2$
coordinate.

The imaginary part of $\Gamma$ is related to the vacuum persistence
probability through the relation:
\begin{equation}
|\langle 0_{\text{out}} | 0_{\text{in}} \rangle|^2
= e^{-2 \, \text{Im} \, \Gamma},
\end{equation}
which represents the probability of no particles being created from the
vacuum field \cite{Schwinger:1951}. The total probability $P$ for the creation
of a fermion pair is expressed as follows:
\begin{equation}
P = 1 - e^{-2 \, \text{Im} \, \Gamma} \simeq 2 \, \text{Im} \, \Gamma,
\end{equation}
where we have considered $\text{Im} \, \Gamma\ll1$. The first significant
process to consider is the formation of a single particle-antiparticle pair.
By retaining only the first non-trivial term in the perturbative expansion of
$\Gamma$, we can calculate the probability of producing just one such pair.

Let us consider the perturbative expansion for $\Gamma$. For clarity, we
will construct the expansion for the case of a single moving wall,
mentioning results for the other cases more succinctly.

\subsection{Single boundary}
We first deal with the case of a single wall, evolving around an average
configuration, which, in our choice of coordinates, is $x^2=0$. 
The effective action for this situation is a functional of $\varphi$,
and we shall use a perturbative approach to obtain the first non-trivial
terms of its  expansion in  powers of that function. We recall that this
function describes the departure of the border from its average
configuration.

To proceed, we write it in an equivalent way by  splitting the action
into two terms, namely,
\begin{equation}\label{eq:defgef1}
e^{i \Gamma(\varphi)} \;=\; 
\langle e^{ i {\mathcal S}_I(\varphi)} \rangle 
\end{equation}
where ${\mathcal S}_I \equiv {\mathcal S}- {\mathcal S}_0$ (${\mathcal S}_0
\equiv {\mathcal S}|_{\varphi=0}$), and:
\begin{equation}
\langle \ldots \rangle \;=\; \frac{\int {\mathcal D}\psi {\mathcal
D}{\bar\psi} \,\ldots\,  e^{i {\mathcal S}_0({\bar\psi},\psi)}}{\int 
{\mathcal D}\psi {\mathcal D}{\bar\psi} \, 
e^{i {\mathcal S}_0({\bar\psi},\psi)}} \;.
\end{equation}
Here,
\begin{equation}
\label{eq:SI}
{\mathcal S}_I \;=\; - \int d^3x \, v(x) \, \bar{\psi}(x) \psi(x) \;,\;\;\;
{\mathcal S}_0(\bar{\psi},\psi) \;\equiv\; {\mathcal S}({\bar\psi},\psi;V_0) 
\end{equation}
with:
\begin{equation}
v(x) \;\equiv\; g \, \left[ \sqrt{1 - \partial_\alpha
\varphi(x_\shortparallel) \partial^\alpha \varphi(x_\shortparallel)} 
\; \delta\big(x^2 - \varphi(x_\shortparallel) \big) 
\,-\, \delta(x^2)  \right] \;.
\end{equation}
We shall expand $\Gamma$ in powers of $\varphi$ up to the second order. It
is rather straightforward to see that, to this order, we need
${\mathcal S}_I$ just to the first order, 
\begin{equation}
{\mathcal S}_I \;=\; g \,  \int d^3x \, \varphi(x_\shortparallel) \,
\delta'(x^2) \, \bar{\psi}(x) \psi(x) \;+\;\ldots
\end{equation}

To evaluate the different terms emerging from the expansion of $\Gamma$ in
powers of the deformation, we need functional averages of fields, with
a weight determined by the quadratic action ${\mathcal S}_0$. By Wick's
theorem,  the resulting terms may be written in terms of the relevant
pairwise contractions of fermion operators, the only non trivial one among
them being the fermion propagator, $S_F$,
\begin{align}\label{eq:defprop}
S_F(x,y) &=\; \langle \psi(x) \bar{\psi}(y) \rangle \nonumber\\
&=\; S_F(x_\shortparallel-y_\shortparallel; x^2, y^2)  \;,
\end{align}
resulting from the action ${\mathcal S}_0$, where we have used the time
independence and $x^1$ translation invariance of $V_0$. 

In the expansion in powers of $\varphi$, the would-be first order term
vanishes, because of our condition on its time average (the term is
proportional to $\int d^2x_\shortparallel \, \varphi(x_\shortparallel)$),
which equals zero. 
To the second-order, we also have real terms which would renormalize the
kinetic term and mass terms of $\varphi$. Discarding them since they do not
contribute to the vacuum decay probability, we are then left with a term,
denoted by $\Gamma_2$, and depending on a kernel $\gamma$ as the only
source of dissipative effects:
\begin{equation}\label{eq:gamma2}
\Gamma \;\simeq \; \Gamma_2 \,\equiv \,
\frac{1}{2} \int d^2x_\shortparallel \int d^2y_\shortparallel \,
\varphi(x_\shortparallel) \,   
\gamma(x_\shortparallel-y_\shortparallel) \,
\varphi(y_\shortparallel) \;,
\end{equation}
where
\begin{equation}
\label{eq:gammacoord}
\gamma(x_\shortparallel-y_\shortparallel) \,=\, - i g^2 \, 
\partial_{x^2} \partial_{y^2} \,\left\{
{\rm tr}\big[S_F(x_\shortparallel-y_\shortparallel;x^2, y^2)
S_F(y_\shortparallel-x_\shortparallel;y^2,
x^2)\big]\right\}\Big|_{x^2,y^2 \to 0} \;.
\end{equation}
Rather than calculating $\gamma(x_\shortparallel-y_\shortparallel)$ directly,
it is convenient to do so for its Fourier transform with respect to the
$x_\shortparallel$ coordinates. Indeed, transforming the propagator and the
deformation:
\begin{align}
S_F(x_\shortparallel-y_\shortparallel; x^2, y^2) &=\, 
\int \frac{d^2p_\shortparallel}{(2\pi)^2} \, e^{-i p_\shortparallel \cdot
(x_\shortparallel -y_\shortparallel)} \, \widetilde{S}_F(p_\shortparallel;
x^2, y^2) \nonumber\\
\varphi(x_\shortparallel) &=\; \int \frac{d^2p_\shortparallel}{(2\pi)^2} \, 
e^{-i p_\shortparallel \cdot x_\shortparallel } \,
\widetilde{\varphi}(p_\shortparallel) \;, 
\end{align}
the expression for $\Gamma_2$ in (\ref{eq:gamma2}) becomes:
\begin{equation}
\label{eq:Gamma2fourier}
\Gamma_2 \;=\; \frac{1}{2} \, \int \frac{d^2k_\shortparallel}{(2\pi)^2} 
\, \widetilde{\gamma}(k_\shortparallel) \, |
\widetilde{\varphi}(k_\shortparallel)|^2  \;,
\end{equation}
where:
\begin{equation}
\label{eq:gammatilde}
\widetilde{\gamma}(k_\shortparallel) = - i g^2
\, \int \frac{d^2p_\shortparallel}{(2\pi)^2}
\, \partial_{x^2} \partial_{y^2} \,
{\rm tr}\big[\widetilde{S}_F(k_\shortparallel + p_\shortparallel;x^2, y^2)
\widetilde{S}_F(k_\shortparallel;y^2, x^2)\big]\Big|_{x^2,y^2 \to 0} \;.
\end{equation}

The next step, in order to evaluate $\widetilde{\gamma}$, involves 
knowing the derivatives of the fermion propagator with respect to the
$x^2, y^2$ coordinates, and then taking the limit when they approach a
static boundary. To that end, we start from the property that the
propagator is the Green's function for the Dirac equation, therefore:
\begin{align}\label{eq:DiracEq}
i \gamma^2 \partial_{x^2} \widetilde{S}_F(p_\shortparallel; x^2, y^2)
\,+\,\left(\not \! p_\shortparallel \,-\, m - g \delta(x^2)\right) \, 
\widetilde{S}_F(p_\shortparallel; x^2, y^2) 
& =\; i \, \delta(x^2 - y^2) \nonumber\\
- i \partial_{y^2}\widetilde{S}_F(p_\shortparallel; x^2, y^2)\gamma^2 
\,+\,\widetilde{S}_F(p_\shortparallel; x^2, y^2) \,
\left(\not \! p_\shortparallel \,-\, m - g \delta(y^2)\right) \,
&=\; i \,\delta(x^2 - y^2) \;.
\end{align}
Hence, we can write the derivatives required in (\ref{eq:gammatilde}) as
follows,
\begin{align}\label{eq:Derivs}
\big[\partial_{x^2} \widetilde{S}_F(p_\shortparallel; x^2,y^2)\big]
\Big|_{x^2, y^2 \to 0} &=\;
- i \gamma^2 \left(\not \! p_\shortparallel \,-\, m \right)
\widetilde{S}_F(p_\shortparallel; 0, 0) \nonumber\\
\big[\partial_{y^2} \widetilde{S}_F(p_\shortparallel; x^2, y^2)\big]
\Big|_{x^2, y^2 \to 0} &=\; 
\widetilde{S}_F(p_\shortparallel;0,0) \left(\not \! p_\shortparallel
\,-\, m \right) i \gamma^2 \nonumber\\
\big[\partial_{x^2} \partial_{y^2} \widetilde{S}_F(p_\shortparallel; x^2,
y^2)\big]\Big|_{x^2, y^2 \to 0} &=\; \gamma^2 \left(\not \! p_\shortparallel
\,-\, m \right) \widetilde{S}_F(p_\shortparallel;0,0)
\left(\not \! p_\shortparallel \,-\, m \right) \gamma^2 \;.
\end{align}
Thus, all the derivatives may be written in terms of a single object:
$\widetilde{S}_F(p_\shortparallel;0,0)$ which is, we recall, the propagator
in the presence of the $x^2 = 0$ wall, approaching the wall on both
arguments. This, in turn, can be obtained from
$\widetilde{S}_F^{(0)}(p_\shortparallel; 0, 0)$, its free-space
(i.e., $g=0$) counterpart, as follows:
\begin{equation}
\widetilde{S}_F(p_\shortparallel; 0, 0) \;=\; 
\frac{\widetilde{S}_F^{(0)}(p_\shortparallel;0,0)}{ 1 + i g
\widetilde{S}_F^{(0)}(p_\shortparallel;0,0)} \;,
\end{equation}
with
\begin{equation}
\widetilde{S}_F^{(0)}(p_\shortparallel; 0, 0) \;=\; \frac{1}{2}\, 
\frac{\not \! p_\shortparallel + m}{\sqrt{ p_\shortparallel^2 - m^2}} \;.
\end{equation}
The square root above should be understood as 
\begin{equation}
\sqrt{p_\shortparallel^2 - m^2} \;=\; \theta(p_\shortparallel^2 -
m^2)\, \sqrt{p_\shortparallel^2 - m^2} \,+\, i \, \theta(m^2 -
p_\shortparallel^2)\, \sqrt{m^2 - p_\shortparallel^2 } \,. 
\end{equation}

Introducing the expressions for the derivatives of the propagator into
(\ref{eq:gammatilde}), discarding again terms which have the form of a
mass or kinetic term counterterm, after a lengthy but otherwise
straightforward calculation, we get:
\begin{equation}\label{eq:pimn1}
\widetilde{\gamma}(k_\shortparallel) \;=\; i g^2 \, \int
\frac{d^2k_\shortparallel}{(2\pi)^2}\, {\rm tr}\big[ 
\gamma^2 \not \! k_\shortparallel  \, \widetilde{S}_F(p_\shortparallel +
k_\shortparallel; 0, 0) \, \gamma^2 \not \! k_\shortparallel
\widetilde{S}_F(p_\shortparallel; 0, 0) \big] \;.
\end{equation}
Let us now reinterpret (\ref{eq:pimn1}) in $1+1$ dimensional terms.
Firstly, we note that the object $\widetilde{S}_F(p_\shortparallel) \equiv
\widetilde{S}_F(p_\shortparallel;0,0)$ acts as the propagator corresponding
to fermionic fields on the boundary, namely, it is the Fourier transform
of
\begin{equation}
S_F(x_\shortparallel - y_\shortparallel;0,0) \;=\; 
\langle \psi(x_\shortparallel,0) \bar{\psi}(y_\shortparallel,0) \rangle \;.
\end{equation}
Besides, (\ref{eq:pimn1}) is a loop integral over a $1+1$ dimensional
momentum. Finally, $\gamma^2$ behaves as a $\gamma^5$ chirality matrix
from the $1+1$ dimensional point of view,  and therefore we have the relation:
\begin{equation}
\gamma^2 \gamma^\alpha \;=\; - i \varepsilon^{\alpha\beta} \gamma_\beta
\;.
\end{equation}
This may be taken advantage of in order to write:
\begin{equation}
\widetilde{\gamma}(k_\shortparallel) \;=\;  
\varepsilon^{\alpha\alpha'}k_\alpha \, \varepsilon^{\beta\beta'} k_\beta \, 
\widetilde{\Pi}_{\alpha'\beta'}(k_\shortparallel)
\end{equation}
where $\widetilde{\Pi}_{\alpha'\beta'}(k_\shortparallel)$ denotes a $1+1$
dimensional version of the {\em electromagnetic\/} vacuum polarization tensor
corresponding to Dirac fields on the boundary: 
\begin{equation}\label{eq:Pi}
\widetilde{\Pi}_{\alpha\beta}(k_\shortparallel) \;\equiv\; - i g^2 \,
\int \frac{d^2p_\shortparallel}{(2\pi)^2} \,
\textrm{tr}\big[ \, \gamma_\alpha \, \widetilde{S}_F(p_\shortparallel +
k_\shortparallel) \, \gamma_\beta \, \widetilde{S}_F(p_\shortparallel) \,
\big]\;.
\end{equation}

The full, $2+1$ dimensional object of which the above is a projection, is:
\begin{align}
\Pi_{\mu\nu}(x,y) &=\; \int \frac{d^2k_\shortparallel}{(2\pi)^2}
e^{- i k_\shortparallel \cdot (x_\shortparallel - y_\shortparallel)} \,
\widetilde{\Pi}_{\mu\nu}(k_\shortparallel;x^2, y^2) \nonumber\\
 &=\; - i g^2 \, 
\textrm{tr}\big[\gamma_\mu \, S_F(x,y) \, \gamma_\nu \, S_F(y,x) \, \big]
\;,
\end{align} 
so that: $\widetilde{\Pi}_{\alpha\beta}(k_\shortparallel) = g_\alpha^\mu
g_\beta^\nu \, [\widetilde{\Pi}_{\mu\nu}(k_\shortparallel;x^2,
y^2)]\big|_{x^2, y^2 \to 0}$, where we used the mixed tensor version of the 
$2+1$ dimensional metric $[g_{\mu\nu}] = {\rm diag}(1,-1,-1)$.

Let us recall that this expression has appeared when we wrote the
derivatives of the propagator with respect to the coordinates normal to the
wall, by using the equation satisfied by the propagator. A related property
is that the reduced tensor does not satisfy a projected version of the Ward
identity in $2+1$ dimensions, namely, $\partial_\mu \Pi^{\mu\nu}(x,y)= 0$,
which holds true, does not imply $k_\shortparallel^\alpha
\widetilde{\Pi}_{\alpha\beta}(k_\shortparallel) = 0$. Indeed, current may
flow from a boundary to the bulk.

We have then arrived to an expression with a vacuum polarization tensor in a
system where the fermions are coupled to an effective electromagnetic field,
with a ``gauge field'' $A_\alpha(x_\shortparallel)$
associated to the deformation, 
\begin{equation}
A^\alpha(x_\shortparallel) \;=\; \varepsilon^{\alpha\beta}
\partial_\beta \varphi(x_\shortparallel) \;,
\end{equation}
such that the second order term may be written, in Fourier or coordinate
space, as follows:
\begin{align}
\Gamma_2 &=\; \frac{1}{2} \int
\frac{d^2k_\shortparallel}{(2\pi)^2} \,
\widetilde{A}^*_\alpha(k_\shortparallel)
\widetilde{\Pi}^{\alpha\beta}(k_\shortparallel)
\widetilde{A}_\beta(k_\shortparallel) \nonumber\\
&= \; \frac{1}{2} \int d^2x_\shortparallel \int d^2y_\shortparallel \,
A_\alpha(x_\shortparallel)
{\Pi}^{\alpha\beta}(x_\shortparallel - y_\shortparallel)
A_\beta(y_\shortparallel) \;.
\end{align}
It is important to recall that the effective gauge field is not free to
undergo gauge transformations, since by construction it has the form of a
$1+1$ dimensional curl of the scalar $\varphi$. So, even though there is no
gauge invariance with respect to gauge transformations of $A_\alpha$, 
 a would-be ``pure gauge'' configuration $A_\alpha = \partial_\alpha
 \omega$ should give no
contribution. That is indeed the fact, since it is tantamount to:
\begin{equation}
\partial^\alpha \omega = \varepsilon^{\alpha\beta} \partial_\beta
\varphi \;. 
\end{equation}
But the only solution to the equations above, by recalling that we are
perturbing in powers of the deformation, which therefore has to be small,
and as a consequence vanish at infinity, is a vanishing deformation.

By analogy with the electromagnetic vacuum polarization, we see that an
imaginary part will appear when the external field (determined by the
deformation) is such that it may create fermion pairs. The imaginary part,
on the other hand, may be computed by using standard techniques and found
exactly  for the case of massless fermions. By computing the momentum integral
in (\ref{eq:Pi}) by dimensional regularization, and taking its imaginary part,
we get:
\begin{equation}
{\rm Im} \, \widetilde{\gamma}(k_\shortparallel) \; = \; \frac{1}{16} \,
\bigg[\frac{\frac{g}{2}}{1 + \big(\frac{g}{2}\big)^2}\bigg]^2 \,
\theta(|k_\shortparallel^0| - |k_\shortparallel^1|) \,
[(k_\shortparallel^0)^2 - (k_\shortparallel^1)^2]^2.
\end{equation}

We observe the dissipative effects are maximized when $g = 2$.
This behavior is consistent with the results for the static Casimir effect,
where the effects are also maximized for the same value of $g$
\cite{Fosco:2008}.

\subsection{Two walls, one of them static}\label{sec:twowalls}
We now consider the case of a system consisting of two walls, $L$ and $R$,
where $L$ undergoes motion, with a position determined by the equation
$x^2 = \varphi_L(x_\shortparallel)$, where $R$ remains static, at $x^2 = a$.
As in the single-wall case, the function $\varphi(x_\shortparallel)$
denotes the departure of the moving wall from its average position, and
satisfies the relation $\int d^2x_\parallel \, \varphi(x_\shortparallel) = 0$.

The potential for this model is (\ref{eq:V_twowalls}), and we enforce bag
boundary conditions on both walls by setting $g_L = g_R = 2$, i.e., we study
the dissipative effects of a cavity whose interior is between $L$ and $R$.
These boundary conditions must be applied as one approaches the walls from
the interior of the cavity.

Expanding the effective action in terms of the departure, and using the
same arguments as before, we arrive at an identical equation for the
effective action up to second-order in terms of a kernel,
\begin{equation}\label{eq:gamma2r}
\Gamma_2 \,\equiv \,
\frac{1}{2} \int d^2x_\shortparallel \int d^2y_\shortparallel \,
\varphi_L(x_\shortparallel) \,   
\gamma_{LL}(x_\shortparallel-y_\shortparallel) \,
\varphi_L(y_\shortparallel) \;,
\end{equation}
with:
\begin{equation}
\gamma_{LL}(x_\shortparallel-y_\shortparallel) \,=\,- i \, g_L^2 \,
\partial_{x^2} \partial_{y^2} \,
{\rm tr}\big[S_F(x_\shortparallel-y_\shortparallel;x^2, y^2)
S_F(y_\shortparallel-x_\shortparallel;y^2, x^2)\big]\Big|_{x^2,y^2 \to 0},
\end{equation}
where the fermion propagator now corresponds to a Dirac field in the
presence of {\em two} (rather than one) static boundaries, and it is
evaluated at the position of the wall $L$.

Again, this may be written in an entirely analogous way as in the single
boundary case, yet with a gauge field defined in terms of $\varphi_L$, and
with a vacuum polarization tensor:
\begin{equation}\label{eq:PiRR}
\widetilde{\Pi}^{LL}_{\alpha\beta}(k_\shortparallel) \;\equiv\; - i g_L^2 \,
\int \frac{d^2p_\shortparallel}{(2\pi)^2} \,
\textrm{tr}\big[ \, \gamma_\alpha \, \widetilde{S}_F(p_\shortparallel +
k_\shortparallel;a,a) \, \gamma_\beta \,
\widetilde{S}_F(p_\shortparallel;a,a) \,
\big]\;.
\end{equation}

In order to compute the imaginary part of $\Gamma$, we adopt a different
approach to the one that we took before. Following the procedure outlined
\cite{OscBag1+1}, we decompose the propagator into positive and negative
energy projectors. These projectors are constructed in terms of the
eigenfunctions of the unperturbed Hamiltonian of the system with two static
walls and $g_L = g_R = 2$. The decomposition of the propagator is as follows:

\begin{equation}
\begin{aligned}
S_F(x_\shortparallel-y_\shortparallel; x^2, y^2) =
\int dk^1 \sum_{k^2}
\big[\theta(x^0 - y^0) \, e^{-i k^0(x^0 - y^0)} \\
\mathcal{P}_\mathbf{k}^+(x^1-y^1;x^2,y^2)
- \, \theta(y^0 - x^0) \, e^{-i k^0(y^0 - x^0)} \,
\mathcal{P}_\mathbf{k}^-(x^1-y^1;x^2,y^2) \big] \; ,
\end{aligned}
\end{equation}

and the energy projectors introduced are expressed as:
\begin{equation}
\begin{aligned}
\mathcal{P}_\mathbf{k}^{+}(x^1-y^1;x^2,y^2) &\;=\;
u_\mathbf{k}(x^1, x^2) \, \bar{u}_\mathbf{k}(y^1, y^2), \\
\mathcal{P}_\mathbf{k}^{-}(x^1-y^1;x^2,y^2) &\;=\;
v_\mathbf{k}(x^1, x^2) \, \bar{v}_\mathbf{k}(y^1, y^2).
\end{aligned}
\end{equation}

The eigenfunctions
$u_\mathbf{k}(x^1,x^2) \equiv \psi_{\mathbf{k},+}(x^1,x^2)$ and
$v_\mathbf{k}(x^1,x^2) \equiv \psi_{\mathbf{k},-}(x^1,x^2)$, are positive-
and negative-normalized solutions of Dirac equation with bag boundary
conditions:
\begin{equation}\label{eq:solutions}
\begin{aligned}
&\psi_{\mathbf{k},\pm}(x^1, x^2) \;=\; 
\sqrt{\ell} \, N_\mathbf{k} \, e^{\pm i k^1 x^1}
\begin{pmatrix}
\pm \left(\frac{k^0 - k^1}{k^2} \right) \, \sin(k^2 x^2) \\ 
\cos(k^2 x^2) + \frac{m}{k^2} \, \sin(k^2 x^2)
\end{pmatrix},\\
&N_\mathbf{k} \equiv k^2 \, \sqrt{\frac{(k^0 + k^1)}
{2\pi \, k^0 \, (a \, ((k^0)^2 - (k^1)^2) + m)}}.
\end{aligned}
\end{equation}
Here, $N_{\mathbf{k}}$ is a normalization constant, and $\ell$ is the
length of the walls in the $x^1$ direction, which must be regarded in the
limit $\ell \to \infty$. The solutions are orthogonal and normalized as:
\begin{equation}
\int_{-\infty}^{\infty} dx^1 \int_0^a dx^2 \,
\psi_{\mathbf{k}^\prime,\sigma^\prime}^\dagger (x^1, x^2) \,
\psi_{\mathbf{k},\sigma}^{\vphantom{\dagger}}(x^1, x^2) =
\delta(\sigma^\prime k^{\prime 1}
- \sigma k^1) \, \delta_{k^{\prime 2},k^2} \,
\delta_{\sigma^\prime,\sigma}.
\end{equation}

We also define $k^0 \equiv \sqrt{(k^1)^2 + (k^2)^2 + m^2}$, and label the
eigenfunctions by $\mathbf{k} = (k^1, k^2)$, where $k^1 \in \mathbb{R}$,
while $k^2$ takes values determined by the transcendental equation:
\begin{equation}
\cos(k^2 a) \, + \, \frac{m}{k^2} \, \sin(k^2 a) \;=\; 0,
\end{equation}
that yields a discrete spectrum~\cite{Mamaev:1980,Bordag:2009}.
The index $\mathbf{k}$ corresponds to the particles' spatial momenta when
the size of the cavity tends to infinity, i.e., when $a \to \infty$.

Given the previous conventions, and using the propagator written in terms of
the energy projectors, we can evaluate the effective action:
\begin{equation}
\Gamma_2 = - \, 4 \int_{k^1,p^1}
\sum_{k^2,p^2}  \Bigr| \big(\bar{u}_\mathbf{k}(0, x^2)
v_\mathbf{p}(0, x^2) \big)^\prime \Bigr|^2_{x^2 \to 0}
\int \frac{d\nu}{2\pi} \, \frac{|\tilde\varphi_L(\nu,k^1 + p^1)|^2}
{\nu - (E_\mathbf{k} + E_\mathbf{p}) + i \varepsilon},
\end{equation}
where the term $i \varepsilon$ is derived from the Fourier integral
representation of the Heaviside function.

Now, by extracting the imaginary part of $\Gamma_2$, we can determine the
vacuum-decay probability:
\begin{align}
\label{eq:Ptotal}
P \simeq 2 \, \textrm{Im} \, \Gamma_2 \;=\;
\int_{k^1, p^1} \sum_{k^2,p^2} \, \rho(\mathbf{k}, \mathbf{p}) \; ,
\end{align}
where we have identified the pair-production probability density:
\begin{align}
\label{eq:ProbDen}
\rho(\mathbf{k}, \mathbf{p})
&= 4 \, \Bigr| \big(\bar{u}_\mathbf{k}(0, x^2)
v_\mathbf{p}(0, x^2) \big)^\prime \Bigr|^2_{x^2 \to 0} \,
|\tilde\varphi_L(k_\shortparallel + p_\shortparallel)|^2 \nonumber\\
&= 4 \, \ell^2 N_{\mathbf{k}}^2 \, N_{\mathbf{p}}^2 \,
\big((k^0 - k^1) - (p^0 - p^1)\big)^2 \,
|\tilde\varphi_L(k_\shortparallel + p_\shortparallel)|^2 \;.
\end{align}

The quantity $\rho(\mathbf{k},\mathbf{p})$ must be understood as a
probability density in the $\mathbf{k}$-space for a wall with length $\ell$.
For further illustration, if we consider rigid motion, i.e., the departure
$\varphi_{L}$ is independent of $x^1$, the expression simplifies to
$|\tilde{\varphi}_L(k_{\shortparallel} + p_{\shortparallel})|^2
= \frac{2\pi}{\ell} \, |\tilde{\varphi}_L(k^0 + p^0, 0)|^2$, which allows us
to write the probability density per unit length of the wall:
\begin{equation}
\frac{\rho(\mathbf{k}, \mathbf{p})}{\ell}
= 8\pi \, N_{\mathbf{k}}^2 \, N_{\mathbf{p}}^2 \,
\big((k^0 - k^1) - (p^0 - p^1)\big)^2 \,
|\tilde\varphi_L(k^0 + p^0, 0)|^2 \;.
\end{equation}

The results obtained here serve to extend the formula derived
in~\cite{OscBag1+1}. In particular, the original formula applicable to
$1+1$ dimensions can be replicated by setting $k^1 = 0$ and $p^1 = 0$, with
the exception of a normalization factor of $\frac{1}{(2\pi)^2}$.
Additionally, we observe that when we interchange the roles of $L$ and $R$,
and use the same function for the departure as before, the result for $P$
remains unchanged. This consistency is due to the invariance of the
unperturbed action under parity transformations.

Finally, we mention that employing the methodologies outlined in
\cite{OscBag1+1}, we can further corroborate our previous findings through
the computation of the matrix elements of the T-matrix:
\begin{equation}
\rho(\mathbf{k}, \mathbf{p}) = |\langle f|T|i\rangle|^2 \;.
\end{equation}
Within this context, the calculation of the transition amplitude considers
initial state as the unperturbed vacuum of the system:
$|i\rangle \equiv |0\rangle$ and a fermion anti-fermion pair in the final
state:
$|f\rangle \equiv b_{\mathbf{k}\vphantom{\mathbf{p}}}^{\dagger}
d_{\mathbf{p}\vphantom{\mathbf{k}}}^{\dagger \vphantom{\dagger}}|0\rangle$,
where $b_\mathbf{k}^\dagger$ and $d_\mathbf{p}^\dagger$ are the particle and
anti-particle creation operators, respectively.

\subsection{Single wall as the boundary of a semi-infinite space}
In this subsection, we explore the scenario involving a single wall at the
boundary of a semi-infinite space.

To study this particular case, we build on the results from subsection
(\ref{sec:twowalls}), where the cavity's width is considered in the limit as
it tends towards infinity ($a \to \infty$). This asymptotic behavior implies
that only the wall denoted by $L$ will remain. By considering this limit,
the vector labels originally described in the solutions (\ref{eq:solutions})
now correspond to the momenta of the particles, with the component in the
$x^2$ direction taking continuous values, $k^2 \in \mathbb{R}_{0}^{+}$.
This shift transforms the discrete sums found in (\ref{eq:Ptotal}) into
integrals, which must be appropriately weighted in the momentum space.

After implementing these modifications, we arrive at the following expression:
\begin{equation}
P = \int_{\mathbf{k}, \mathbf{p}} \, \rho(\mathbf{k}, \mathbf{p}) \; ,
\end{equation}
with:
\begin{equation}
\rho(\mathbf{k},\mathbf{p}) = \frac{4 \ell^2}{(2\pi)^4} \,
\frac{(k^2)^2 (p^2)^2 \left((k^0 - k^1) - (p^0 - p^1)\right)^2}
{k^0 \, p^0 \, (k^0 - k^1) \, (p^0 - p^1)} \,
|\tilde\varphi_{L}(k_\shortparallel + p_\shortparallel)|^2 \; .
\end{equation}

\subsection{Two moving walls}
We conclude this section by writing the result for two moving walls.The wall
denoted as $L$ possesses an average position of $x^2 = 0$ and a deviation of
$\varphi_L(x_\shortparallel)$. Similarly, the wall denoted as $R$ holds an
average position of $x^2 = a$ and deviates by $\varphi_R(x_\shortparallel)$.
Again, we enforce bag boundary conditions on both walls: $g_L = g_R = 2$.
Consequently, the resultant effective action to the second order comprises
a sum of three terms; two relate to a static wall and a moving one, plus a
mixed term:
\begin{equation}
\Gamma_2 = \Gamma_2^L + \Gamma_2^R + \Gamma_2^{LR}.
\end{equation}
The mixed term, denoted by $\Gamma_2^{LR}$, represents the interaction
between the walls, and can be expressed as:
\begin{align}
\Gamma_2^{LR} &=\; \int d^2x_\shortparallel \int d^2y_\shortparallel \,
\varphi_L(x_\shortparallel) \, \gamma_{LR}(x_\shortparallel
- y_\shortparallel) \, \varphi_R(y_\shortparallel) \nonumber\\
& = \; \int d^2x_\shortparallel \int d^2y_\shortparallel \,
A_L^\alpha(x_\shortparallel) \, \Pi^{LR}_{\alpha\beta}(x_\shortparallel
- y_\shortparallel) \, A_R^\beta(y_\shortparallel)\;,
\end{align}
with
\begin{align}
\gamma_{LR}(x_\shortparallel - y_\shortparallel) = 
- i \, g_L g_R \, \partial_{x^2} \partial_{y^2} \,
{\rm tr}\big[S_F(x_\shortparallel-y_\shortparallel;x^2, y^2) \nonumber\\
S_F(y_\shortparallel-x_\shortparallel;y^2, x^2)\big]
\Big|_{x^2 \to 0, y^2 \to a}\;,
\end{align}
and
\begin{equation}\label{eq:PiLR}
\widetilde{\Pi}^{LR}_{\alpha\beta}(k_\shortparallel) \;\equiv\;
- i g_L g_R \, \int \frac{d^2p_\shortparallel}{(2\pi)^2} \,
\textrm{tr}\big[ \, \gamma_\alpha \, \widetilde{S}_F(p_\shortparallel
+ k_\shortparallel;0,a) \, \gamma_\beta \,
\widetilde{S}_F(p_\shortparallel;a,0) \,
\big]\;.
\end{equation}

Just as with the effective action, the probability density function comprises
three elements:
\begin{equation}
	\rho = \rho_L + \rho_R + \rho_{LR},
\end{equation}
where
\begin{align}
\rho_{LR}(\mathbf{k},\mathbf{p}) \;=\; &2 \, g_L g_R \, \text{Re} \big[
\, \varphi_{L}^{*}(k_\shortparallel + p_\shortparallel)
\varphi_R(k_\shortparallel + p_\shortparallel) \,
f(\mathbf{k},\mathbf{p})\big],
\end{align}
and
\begin{align}
f(\mathbf{k},\mathbf{p}) \;=\; &
\partial_{x^2} \partial_{y^2} \left[
\left(\bar{u}_{\mathbf{k}}(x^2) v_{\mathbf{p}}(x^2)\right)^{*}
\left(\bar{u}_{\mathbf{k}}(y^2) v_{\mathbf{p}}(y^2)\right)
\right] \Big|_{x^2 \to 0, y^2 \to a} \nonumber\\
= \, & \, \ell^2 N_{\mathbf{k}}^2 \, N_{\mathbf{p}}^2
\left((k^0 - k^1) - (p^0 - p^1)\right) \, \frac{\sin(a k^2)}{k^2}
\frac{\sin(a p^2)}{p^2} \, \nonumber\\
&\left[(k^0 - k^1)\left((p^2)^2 + m^2 \right)
- (p^0 - p^1)\left((k^2)^2 + m^2\right)\right] \, .
\end{align}

From this result, we see that, depending on the values of $\mathbf{k}$ and
$\mathbf{p}$, $\rho_{LR}$'s contribution to the overall pair-production
probability can be either positive or negative. This is particularly
noticeable in the massless case ($m = 0$). In this scenario, the spectrum is
set as $k^2 = (k + \frac{1}{2}) \frac{\pi}{a}$, where
$k = 0, 1, \ldots$, which implies that $\sin(a k^2) = (-1)^{k}$.
Consequently, the preceding equation takes the form:
\begin{align}
f(\mathbf{k}, \mathbf{p}) =
(-1)^{k + p} \, \frac{\ell^{2}}{(2\pi a)^2} \, \frac{k^2 \, p^2}{k^0 \, p^0}
\left((k^0 - k^1) - (p^0 - p^1)\right)
\left[\frac{(p^2)^2}{p^0 - p^1} - \frac{(k^2)^2}{k^0 - k^1}\right] \, .
\end{align}

%%%%%%%%%%%%%%%%%%%%%%%%%%%%%%%%%%%%%%%%%%%%%%%%%%%%%%%%%%%%%%%%%%%%%%%%%%%%%%
%%%%%%%%%%%%%%%%%%%%%%%%%%%%%%%%%%%%%%%%%%%%%%%%%%%%%%%%%%%%%%%%%%%%%%%%%%%%%%
%%%%%%%%%%%%%%%%%%%%%%%%%%%%%%%%%%%%%%%%%%%%%%%%%%%%%%%%%%%%%%%%%%%%%%%%%%%%%%
%%%%%%%%%%%%%%%%%%%%%%%%%%%%%   Conclusions   %%%%%%%%%%%%%%%%%%%%%%%%%%%%%%%%
%%%%%%%%%%%%%%%%%%%%%%%%%%%%%%%%%%%%%%%%%%%%%%%%%%%%%%%%%%%%%%%%%%%%%%%%%%%%%%
%%%%%%%%%%%%%%%%%%%%%%%%%%%%%%%%%%%%%%%%%%%%%%%%%%%%%%%%%%%%%%%%%%%%%%%%%%%%%%
%%%%%%%%%%%%%%%%%%%%%%%%%%%%%%%%%%%%%%%%%%%%%%%%%%%%%%%%%%%%%%%%%%%%%%%%%%%%%%
\section{Conclusions}\label{sec:conc}
In this paper, we studied the effect of fermion pair creation due to the
motion of the boundaries for a system consisting of Dirac fermions in  2+1
dimensions.

The boundary conditions where imposed through a Dirac $\delta$ potential
which, depending on the value of a dimensional parameter, produces different
conditions on the normal component of the fermionic current.
We have shown that the pair creation phenomenon was maximized for a
specific value of the parameter, which correspond to bag boundary conditions.

The study was conducted by the computation of the effective action of the
system. We shown that the effective action of the system can be written
in terms of a vacuum polarization tensor and a gauge field associated with
the deformation of the moving walls. Furthermore, the polarization tensor
itself can be written in terms of the correlation function of two fermion
currents evaluated at the moving boundaries, and computable using the
system's propagators in the presence of static walls.

This work also generalizes the previous findings for the $1+1$ dimensional
case, which were obtained just for cases where there was no current flow through the
boundaries, i.e., when the parameter in the potential took the value $g =
2$. Besides,  we have also considered the situation of having either one or
two walls in motion, showing that a physically consistent picture emerges
in the case of a cavity where one of the walls moves and the position of
one of the other is assumed to be very far away.

%%%%%%%%%%%%%%%%%%%%%%%%%%%%%%%%%%%%%%%%%%%%%%%%%%%%%%%%%%%%%%%%%%%%%%%%%%%%%%
%%%%%%%%%%%%%%%%%%%%%%%%%%%%%%%%%%%%%%%%%%%%%%%%%%%%%%%%%%%%%%%%%%%%%%%%%%%%%%
%%%%%%%%%%%%%%%%%%%%%%%%%%%%  Acknowledgements %%%%%%%%%%%%%%%%%%%%%%%%%%%%%%%
%%%%%%%%%%%%%%%%%%%%%%%%%%%%%%%%%%%%%%%%%%%%%%%%%%%%%%%%%%%%%%%%%%%%%%%%%%%%%%
%%%%%%%%%%%%%%%%%%%%%%%%%%%%%%%%%%%%%%%%%%%%%%%%%%%%%%%%%%%%%%%%%%%%%%%%%%%%%%
\section*{Acknowledgements}
The authors thank ANPCyT, CONICET and UNCuyo for financial support.

\end{document}